**Direct observation of the cell-wall remodeling in adhering *Staphylococcus aureus 27217*: an AFM study supported by SEM and TEM.**

Rym Boudjemaa,[a] Karine Steenkeste[a], Alexis Canette [b, c], Romain Briandet,[b] Marie-Pierre Fontaine-Aupart,[a] Christian Marlière[a].

[a] Institut des Sciences Moléculaires d'Orsay (ISMO), CNRS, Université Paris-Sud, Université Paris-Saclay, Orsay, France;
[b] Micalis Institute, INRA, AgroParisTech, Université Paris-Saclay, Jouy-en-Josas, France ;
[c] Institut de Biologie Paris-Seine (FR 3631) ; Unité Mixte de Service (UMS 30) d'Imagerie et de Cytométrie (LUMIC) ; Sorbonne Université, CNRS, Paris, France.

## Abstract

We took benefit from Atomic Force Microscopy (AFM) in the force spectroscopy mode to describe the time evolution –over 24h- of the surface nanotopography and mechanical properties of the strain Staphylococcus aureus 27217 from bacterial adhesion to the first stage of biofilm genesis. In addition, Transmission Electron Microscopy (TEM) and Scanning Electron Microscopy (SEM) experiments allowed identifying two types of self-adhering subpopulations (the so-called "bald" and "hairy" cells) and revealed changes in their relative populations with the bacterial culture age and the protocol of preparation. We indeed observed a dramatic evanescing of the "hairy" subpopulation for samples that underwent centrifugation and resuspension processes. When examined by AFM, the "hairy" cell surface resembled to a herringbone structure characterized by upper structural units with lateral dimensions of ~70 nm and a high Young modulus value (~ 2.3 MPa), a mean depth of the trough between them of ~15nm and a resulting roughness of ~5nm. By contrast, the "bald" cells appeared much softer (~ 0.35 MPa) with a roughness one order of magnitude lower. We observed too the gradual detachment of the herringbone patterns from the "hairy" bacterial envelope of cell harvested from a 16h old culture and their progressive accumulation between the bacteria in the form of globular clusters. The secretion of a soft extracellular polymeric substance was also identified that, in addition to the globular clusters, may contribute to the initiation of the biofilm spatial organization.

## Keywords

**Introduction**

*Staphylococcus aureus* (*S. aureus*) is a Gram-positive pathogen implicated in a wide range of hospital-acquired infections and often associated with biofilm formation on medical implants[1]. Biofilm growth involves bacterial adhesion to (a)biotic surfaces, followed by cell-cell interactions leading to microcolonies and further mature 3D structure, encased in a self-secreted exopolymeric matrix[2]. Initial bacterial adhesion is associated with changes in cellular physiology, enabling bacteria to cope with antimicrobials aggression and thus rendering associated infections very difficult to treat[3].

Among the key factors promoting bacterial adhesion, cell wall constituents (proteinaceous adhesins, capsular polysaccharides, ….) and cell heterogeneity fulfill an important role in substrate and intercellular interactions[1–6]. Numerous studies in the literature provided interesting information on *S. aureus* cell wall morphology, morphogenesis and degradation in connection with treatments, resistance and virulence by the use of electron microscopies (Scanning Electron Microscopy, SEM and Transmission Electron Microscopy, TEM)[7,8]. The development of cryomethods permitted, for this specie, better ultrastructural imaging of, for instance, the extracellular matrix[9] and the identification of news structures, *e.g.* those in the cell external layers during division[10]. Atomic Force Microscopy allowed improving this exploration by giving access, at the nanoscale, to both morphogenesis and mechanical properties of adhering living bacteria in their native liquid environment[11–22]. The force spectroscopy mode associated or not with biospecific probes has proved to be very useful for the quantification of subcellular chemical heterogeneities, but also for the characterization of bacterial interactions with each other, with immune cells or with specific molecules such as lectins, antimicrobials, antibodies, …[23–27] Notably, recent studies evaluated some of *S. aureus* surface decorations that mediate cell-cell interactions (PIA, SraP, SdrC, SasG, …), revealing electrostatic and/or specific homophilic binding between proteins of the interacting cells[4,24,28,29]. However, little is currently known about the bacterial cell wall remodeling with time, from the early growth phase to the initiation of biofilm formation. It was shown that the cell wall components can be either covalently anchored *to S. aureus* cell wall or loosely-attached to mediate cell-surface and cell-cell adhesions[4,30]. In this latter case, it cannot be excluded that the surface decorations might be removed from the cell wall during "harsh" sample preparation, (centrifugation, resuspension, filtration, …).

In this context, this study took benefit from AFM experiments in the approach-retract scanning (also called force-spectroscopy) mode, supported by TEM and SEM experiments, to describe the evolution of the surface nanotopography of the strain *S. aureus* ATCC 27217 with its growth state by taking special care to preserve the integrity of the native bacterial cell wall during sample conditioning.

**Material and Methods.**

**Bacterial strain and growth condition.**
The strain *S. aureus* ATCC 27217 was used in this study, a methicillin-susceptible reference strain that is able to form biofilms both *in vitro* and *in vivo*[31,32]. The strain was stored at –80°C in Trypticase Soy Broth (TSB, bioMérieux, France) containing 20% (vol/vol) glycerol. Frozen cells were cultured in TSB at 37°C without agitation and then harvested at different times of bacterial growth upon need. The planktonic bacterial suspensions were used as such (so called "non-centrifuged") or after being pelleted and re-suspended (so called "centrifuged").

**Scanning Electron Microscopy (SEM).**

In order to preserve cell wall structure, the "non-centrifuged" planktonic bacterial suspensions were directly fixed for 10 min at room temperature in distilled water containing 4% (v/v) glutaraldehyde (2% glutaraldehyde final concentration). For "centrifuged" samples, the planktonic cell suspensions were pelleted at 3000$g$ for 10 min at 4°C before fixation in 0.10 M cacodylate buffer containing 2% (v/v) glutaraldehyde (pH 7.2).

Glutaraldehyde is a specific protein fixator[33] that creates relatively rapid (around 0.5mm/h rate of penetration at room temperature) intra and intermolecular irreversible crosslinking between the amino groups of proteins. This results in the structure stabilization of the cell membrane (by the preservation of the membrane proteins) and its surface appendages by the preservation of the proteinaceous and glyco-proteinaceous compounds of the cell wall[34].

A 40 µL drop of fixed cells was deposited onto a sterile aluminum coupon (10-mm diameter, sterilized just before use by sonication in ethanol and dried during UV exposure) placed into one well of a 24 well polystyrene plate. Sedimentation of bacteria lasted 1.5 h at room temperature. Samples were fixed again via careful immersion in a 0.10 M cacodylate buffer containing 2.5% (v/v) glutaraldehyde (pH 7.2) for 10 min at room temperature, followed by overnight waiting time at 4°C. Samples were then washed three times for 5 min with 0.1 M sodium cacodylate buffer. Thereafter cells were dehydrated in an ethanol series (30 %, 50 %, 70 %, 90 % v/v with distilled water, and 3 times with 100 % ethanol, 10 min for each step). Samples were critical-point dried (Quorum Technologies K850, Elexience, France) at 70 bar and 37°C with liquid $CO_2$ as the transition fluid and then depressurized slowly (400 $cm^3 \cdot min^{-1}$). Each aluminum support carrying the sample was then mounted on an aluminum stub with double-sided carbon tape. Samples were sputter-coated (Polaron SC7640, Elexience, France) in Ar plasma with Pt at 10 mA and 0.8 kV over duration of 200 s.

Observations were performed in a field-emission SEM (Hitachi S4500, Japan) in high vacuum, with a secondary electron low detector, at 2 kV and 16mm working distance, at the MIMA2 imaging platform (INRA Jouy-en-Josas, www6.jouy.inra.fr/mima2/).

**SEM Image treatment**

We developed a systematic approach to extract from SEM images the ratio of the "hairy" to the "bald" bacteria (see below in the "results and discussion" section) for the different samples. The bacterial surface structures, if present, are numerically amplified and quantified by a home-made Matlab program[35] (see figure SI1 and its caption for more details) that automatically counts the ratio of the 'on' pixels, where these surface structures are revealed, to the total number of pixels for each bacterium. When this "pixel-ratio" is lower (respectively higher) than 0.5, the case with few surface decoration, the bacterium is said to be of the "bald" ("hairy" resp.) type. As shown below, the value for this chosen threshold (0.5) is fully justified by the fact that the distribution of bacteria is bimodal with maxima centred at pixel-ratios of around 0.30 and 0.75. It must be emphasized that the image processing code – and the different thresholds we used- was kept invariant throughout all the treated SEM images (at a magnification of $10.10^3$ and $20.10^3$).

**Transmission Electron Microscopy (TEM).**

"Centrifuged" and "non-centrifuged" bacteria were prepared and fixed as described previously for SEM observations. Fixed bacteria were kept overnight at 4 °C in a 0.10 M cacodylate and 0.20 M sucrose buffer. They were then washed one time during 5 min with 0.10 M cacodylate buffer, contrasted during 1h with 0.5% OTE in 0.10 M cacodylate buffer, and washed 2 times during 5 min with 0.10 M cacodylate buffer. Samples were post-fixed for 1 h at room temperature in 0.10 M cacodylate buffer containing 1 % (v/v) osmium tetroxide with 1.5% potassium cyanoferrate, and washed twice for 5 min with distilled water. Thereafter cells were dehydrated in an ethanol series (30 %, 50 %, 70 %, 90 % v/v with distilled water, and 3 times with 100 % ethanol, 10 min for each step, except overnight for 70%). A 10 min intermediate bath in propylene oxide was performed. Then, bacteria were impregnated at room temperature in successive mixes of propylene oxide and epon (2:1; 1:1 and 1:2, for 2 h each step), then in pure epon overnight and finally in vacuum conditions. A final inclusion bath with pure epon and DMAE (accelerator) was performed and polymerization was allowed by incubating for 48 h at 60 °C. Ultrathin sections of 70 nm were cut with an ultramicrotome (UC6, Leica, Germany) and deposited on 200 mesh copper platinum grids. Sections were stained for 2 min in Reynolds lead citrate and rinsed in distilled water. Observations were performed using an HT7700 transmission electron microscope (Hitachi, Japan) equipped with an 8 million pixels format CCD camera driven by the image capture engine software AMT, version 6.02, at the MIMA2 imaging platform (INRA Jouy-en-Josas, www6.jouy.inra.fr/mima2/). Images were made at 80 kV in high contrast mode with an objective aperture adjusted for each sample and magnification.

**AFM experiments.**

**(i) Bacterial self-immobilization.**

One fundamental requirement when using AFM is to avoid the sweeping away of bacteria from the scanned region by the AFM tip because of its lateral interactions with the poorly adhered bacteria. To cope with this limitation, several approaches have been employed to immobilize or fix cells (mechanical trapping into membrane pores, chemical coating of substrates,…)[36–38]. Unfortunately, such strategies may induce stressful conditions but, more importantly, alter the bacterial cell physiology and bias the observations. To obtain the most realistic representation, we therefore imaged living bacteria spontaneously adhering on the substratum (RBS and $CaCO_3$-cleaned Indium-Tin Oxide (ITO) glass slides). The harvested planktonic bacterial suspensions were used as such ("non-centrifuged") or after being centrifuged as previously described. 500 µL portions of the "centrifuged" or "non-centrifuged" bacterial cultures were deposited on the cleaned ITO for 1h30 at 37°C. Samples were then both rinsed and refilled with sterile aqueous NaCl (9 g/L) solution supplemented with {$CaCl_2$, $2H_2O$} (50 mg/L). All experiments were performed with, at least, ten cells from three different bacterial cultures.

**(ii) AFM data.**

Atomic force microscopy studies were carried out using a Nanowizard III (JPK Instruments AG, Berlin, Germany) and its electrochemical cell (ECCell® from JPK). The AFM head was working on a commercial inverted microscope (Axio Observer.Z1, Carl Zeiss, Göttingen, Germany). This combined AFM/optical microscope was placed on an isolation vibration table. AFM measurements were performed using a fast-speed approach/retract mode (Quantitative Imaging® (QI) mode, JPK) giving the ability of performing local mechanical properties of a sample (Young's moduli) in the so-called force spectroscopy mode. Force curves were acquired over 128 pixels x 128 pixels images, with a maximum applied force of 1.2 nN for all conditions at a constant approach/retract speed of 150 µm/s (z-range of 500nm). Standard beam AFM probes (CSC38 MikroMasch, NanoAndMore GmbH, Wetzlar, Germany) were used with a nominal value of stiffness around 0.03 N.m$^{-1}$ (precisely measured by thermal noise). This rather low value was chosen to be fitted to the mechanical characteristics of the cell wall and surface appendices. The sensitivity of detection of the vertical deflection thanks to the photodiode system was measured during the approach to a clean glass substrate. Raw data treatment was then performed using home-made Matlab programs and Origin Pro software. Young's moduli were calculated using the Hertz model[39–41].

## Results and Discussion

**Ultrastructure of *S. aureus* cell wall by combining SEM and TEM.**

For SEM and TEM images, *S.aureus* bacteria were harvested after 3h of culture (exponential growing phase) or 18h (stationary phase). For both centrifuged and non-centrifuged samples, and whatever the bacterial growing time, two types of cell surfaces topography were observed: one showing superficial rough structure (so called "hairy" cell) and the other one displaying regular and smooth surface (so called "bald cell") (figure 1). We developed a systematic study of the influence of the culture age and centrifugation on the ratio of the hairy to the bald bacteria from the SEM images (see "material and methods" section).

The related four curves are plotted in figure 2. One common feature is that the distribution is bimodal with maxima centred at a pixel-ratio of 0.30±0.07 (corresponding to the "bald" bacteria) and 0.75±0.07 ("hairy" bacteria) respectively. The first important result is that centrifuged populations (blue and black curves in figure 2) are mainly from the "bald" type: the hairy/bald ratio is around 0.2 whatever the harvesting time. By contrast, for non-centrifuged samples, this ratio is inversed and depends on the harvesting time: the hairy/bald ratio is around 2 for a harvesting time of 3h and it decreases to 0.8 after 18h of culture. It must be noted that hairy and bald bacteria are both observed for cells in or out of the division process. Example of the former case is clearly visible in the SEM images at a much higher magnification of the zone around the septal ring when in formation (figures 1H1, 1B1) or in a slightly higher degree of separation (figure SI2; non-centrifuged cells: left column, centrifuged cells: right column). In this case we note the presence, along the septal line, of structures similar to "holes" as those observed by Touhami *et al.*[42] by AFM in liquid environment or by SEM as in work of Zhou *et al.*[43]

An interesting feature is that this hairy component seems to be a specificity of *S. aureus* strains able to form dense biofilms like the ATCC 27127 strain: the analysis by SEM and TEM of two others reference isolates, widely used for antibiotic testings, namely the methicilin sensitive *S. aureus* strain Newman and the methicilin resistant *S. aureus* strain JE2 (also named USA 300), did not reveal any "hairy" populations in the same experimental conditions and visualisation protocols (data not shown).

These two important observations, firstly, the removal of bacterial surface decorations with a pelleting treatment and, secondly, the evolution of the repartition between both populations of hairy or bald bacteria with the maturation of bacterial culture have been confirmed and enriched by real-time and *in-situ* AFM experiments.

**Direct observation of *S. aureus* bacterial surface by *in-situ* AFM**

For samples harvested at 3h culture, AFM images at a high magnification -(0.4 x 0.4 µm)² scanning areas-, on the top of bacteria, are reported in figure 3. At this resolution, the "hairy" surface appears as a regular distribution of three-dimensional herring-bone patterns (Figure 3.a.1), exhibiting lateral dimensions varying between 50 and 100 nm and a mean depth of the trough between them varying between 10 and 15 nm and a roughness of 4.7 nm RMS (for a (400nm)² area). It must be emphasized that such structural observations have never been reported for *S. aureus* species. On the contrary, the bacteria appeared much smoother (figure 3.a.2) for the centrifuged samples (1.8 nm RMS for a (400nm)² area).

Elasticity maps determined from AFM measurements in approach-retract mode are reported in figures 3.b.1 and 3.b.2. The corresponding histograms depicting the Young's moduli distribution over each image are represented in figures 3c.1 and 3.c.2. For the non-centrifuged samples, three major values (center of Gaussian peaks) were pointed at 0.35 ± 0.03 MPa, 0.95 ± 0.07 MPa and 2.3 ± 0.3 MPa (mean ± SD, $n$ = 32 768 curves from two different cells from two different cultures) with respective contributions (areas under the related peak) of 17 ± 6%, 30 ± 5% and 53 ± 9%. Correlation between structural and elasticity properties indicates that the stiffest component (~ 2.3 MPa) corresponds to the upper (herringbone) patterns of the cell surface while the softer one (~ 0.35 MPa) is related to the deeper valleys, thus likely attributed to the bacterial cell wall. By contrast, the centrifuged cells (3h culture) have a softer and more homogenous surface (major peak in Young's modulus histogram centered at 0.3 MPa, figure 3.c.2) when compared to the raw samples, confirming that the cells decorations with a high value for Young's modulus were removed by pelleting and thus loosely-attached to the cell wall.

Some stages of bacterial division for samples harvested at 3h culture could also be monitored by real-time AFM scanning. In line with electron microscopy observations[10,42,44], the gap between two dividing cells evolution progressed with time (Figure 4a): during a time interval of 35 minutes, both space and depth along the bacterial septum increased from 100 to 160 nm and 50 to 200 nm respectively (figure 4.c). The mean speed leading to cell division (figure 4.c-d) was estimated to 1.5 nm.mn$^{-1}$. Remarkably, 0.4 x 0.4 µm² height scans performed on the two daughter cells from hairy mother-bacteria showed the same herring-bone patterns (figure 4.b) as observed on the top of the dividing cells, strongly suggesting that this over-structure is produced during cell division.

The evolution of the hairy/bald ratio with culture age, already highlighted by SEM study, was further analyzed by AFM. Figure 5 illustrates the evolution of the structural and mechanical properties of "hairy" *S. aureus* bacterial surface with time: the living bacteria, from non-centrifuged samples, were harvested from 3-h to 24-h cultures (figure 5). AFM height images (figures 5.a and 5.b) and elasticity maps (figures 5.c) obtained on the cell top at 3-, 16- and 18-h cultures, always show herring-bone patterns with similar mechanical properties (with a Young's modulus contribution at ~ 2.3 MPa, figures 5.c.1 and 5.d.1). In addition, multiple 70-nm large and 10-20-nm deep holes (figure 5.b.2) were observed, yielding a low value (0.4 ± 0.2 MPa) for elasticity (figure 5.c.2 and 5.d.2). With samples grown for 18 h, this zone of low Young's moduli values appeared in a large band on the cell surface (figure 5.b.3), strongly suggesting a gradual removal of the stiff extracellular decorations over successive patches. This was mainly observed along the septal line in formation (figure 5.a.2, black square). From the acquisition time for an AFM image (~4 minutes in our conditions) we can roughly estimate the typical time for the removal of one scale to ~240s. This gives a rough approximation for the propagation speed of the detachment of cell surface decoration along the peripheral ring of 15nm/mn. The presence of such holes and their coalescence were already evidenced by electron microscopy[7,45]: they were attributed to "the *murosomes*-mediated punching of holes into the peripheral wall for cell"[46], "leaving behind characteristic clefts on the cell surface"[7]. This phenomenon was also observed by AFM experiments on hydrated samples[42]. In this last paper the diameter of the holes was estimated to 50-60nm, similar to our estimation (around 70nm). These results were confirmed by recent SEM experiments[43] revealing "structures in a subpopulation of cells that appeared to be perforation-like holes and cracks along the peripheral ring" as noted in Touhami *et al* paper[42] . The authors interpreted these perforations as "points of mechanical failure that could initiate a propagating crack" which would be responsible of a "popping effect" for the generation of two daughter cells.

For the advanced phase cells growth (24h in culture), bacteria were no longer surrounded by the 2.3 MPa stiff decorations, as observed on the top of the cells (figure 5.c.4). Instead, consistently with earlier AFM studies on *S. aureus*[24,47], they displayed a homogenously smooth (figure 5.b.4) and soft cell surface (figure 5.c.4).

The Young's moduli histograms (areas under peak for each of the three Gaussian peaks) depicted in figures 5.d.1-4, confirm that the contribution of the stiffer component corresponding to the hearing-bone patterns drastically decreases for cell growth at 3h to 20h (53 to 0%). Inversely, the softer component (0.4 MPa), likely related to the naked *S. aureus* cell wall (the peptidoglycan network without its native surface structure), significantly increases from 17 to 90 %. These lower values for Young's modulus were the only one we measured for centrifuged bacteria independently of the growth phase (see figures 3.b-c.2). The intermediate softer value (centred on 0.8MPa) also disappears along time and corresponds to an intermediate zone of the cell surface between the herringbone pattern and the naked *S. aureus* cell wall. These two peaks at lower values for Young's modulus are in agreement with previous data reported in literature[48]. Taking together these electron microscopy and AFM results, we demonstrated that the surface structures of the hairy bacteria were loosely attached to the cell wall as they are removed by centrifugation generating a shaved "bald" bacterial population. It must be emphasized the tendency for *S. aureus* cells to lose their surface decorations -the herringbone motifs- with the age of the culture: the ratio of "hairy" to "bald" bacteria significantly decreases from the early growth to the late ones. This likely explains why this high value of Young's modulus (~ 2.3 MPa) has not been previously reported in studies where the bacterial samples were systematically centrifuged or studied in late stationary phase[28,48]: (Young's modulus ~ 0.35-0.8 MPa).

Another interesting feature was that, besides of the classical round-shape bacteria, globular clusters have been evidence in this late phase (figure 5.a.4, red arrows). Strikingly, we found by focusing on such globular clusters (figure 5.b-c.5) that they revealed similar structural and mechanical properties to those of the surface of hairy bacteria at 3h of growth phase: herringbone patterns (~ 15-nm high) with a roughness of 2.6nm RMS for an 300nm² area (figure 5.b.5), highly stiff (major peak in histogram –figure 5.d.5- of Young's moduli centered at 2.68 ± 0.17 MPa, figure 5.c.5). In view of these results, it is then tempting to speculate that during aging, bacteria get rid of their stiff envelope layer, which further accumulates into globular clusters between cells, likely favouring cell-cell adhesion and further biofilm formation. The question arises on the nature of this stiff component. As discussed above, its mechanical properties discard a soft material like extracellular polysaccharides. A preliminary proteomic analysis[49] of supernatants from centrifugations of bacterial suspensions at different cell growth allowed to identify several candidate proteins known to promote cell-cell interaction during biofilm development, including fibrinogen-binding proteins, immunoglobulin-binding proteins, peptidoglycan-binding proteins, serine-rich proteins or penicillin-binding protein[50–54]. To support the hypothesis of a proteinaceous nature of these decorations we can refer to the role of glutaraldehyde used for SEM experiments, as it acts specifically on the cell surface proteins (see Materials and Methods section). However, the formal identification of these decorations will require further extensive genetic and biochemical analysis.

In addition to this stiff envelope, at 24h, a plume of an extracellular material was shown to be slowly secreted by bacteria as it can be seen in figure 5.a.3 (the bright region inside the blue circle). Such a secretion process was investigated thanks to successive AFM images at a higher magnification (figure 6.a-b). The time evolution of the contour of the plume (color lines in figure 6.c) leads to a rough estimation of the diffusion/secretion coefficient of $20\pm5$ nm²/s (figure 6.d). Elsewhere, the Young's modulus of this extracellular material was estimated to $50\pm10$ kPa, the signature of a very soft material. In a previous AFM study[24], it has been shown that *S. aureus* can produce an extracellular soft layer that drastically increases the softness of the cell surface (~ 45 kPa *vs*. ~ 500 kPa); a finding attributed to the secretion of Polysaccharide Intercellular Adhesin (PIA), a biofilm matrix component. We hypothesize that the observed plume could be related to the secretion of such an extracellular polymeric substance that is also contributing to the completion of the biofilm spatial organization.

**Conclusion.**

The results of this study can be resumed by the scheme of Figure 7. In the exponential growth phase, the cell surface of "hairy" bacteria is surrounded by a stiff extracellular layer resembling herring-bone patterns. Along with phase aging, these patterns get gradually removed by patches, firstly forming holes, secondly zipper-like patterns and finally accumulate into globular clusters sticking between bacteria. During this late stage, bacteria also secrete a very soft extracellular polymeric material, probably contributing to the matrix construction.

Our soft mode of bacterial preparation (without initial centrifugation) to conduct experiments combining SEM, TEM and AFM in the force spectroscopy mode, enabled to analyze in real time the evolution of the structural and mechanical properties of the strain *S. aureus* 27217 in different physiological states. We report for early stage cell population a stiff, 15 nm in thickness, extracellular layer covering the cells, probably of protein composition that progressively detaches with time (form 3h to 18h cell growth) to finally (20 h) agglomerates into globular deposits sticking between the cells. Furthermore, it was also revealed the appearance at about 20-24h, of a soft material as expected for polysaccharide substances, cementing cell agglomerates. Both these stiff and soft structures likely play a crucial role in intercellular adhesion and aggregation and subsequent biofilm formation, a specific character of the *S. aureus* ATCC 27217 bacterial strain. The prevalence of such cell wall decoration within *S. aureus* isolates, as well as their precise nature and function will require extensive additional work.

**Figure captions**

Figure 1:

Observation by SEM and TEM of *S. aureus* ATCC 27217 bacterial strain. Evidence of two types of self-adhering subpopulations: the so-called "*bald*" and "*hairy*" cells. Scale bars: Figure 1.A: 6µm; figures 1.H.1 and 1.B.1: 600nm; ; figures 1.H.3 and 1.B.3: 200nm; figures 1.H.2, 1.H.4, 1.B.2 and 1.B.4: 100nm.

Figure 2:

Variation of the number of bacteria with the value of ratio – the "pixel-ratio"- of the 'on' pixels, where the "hairy" surface structures are revealed by SEM experiments, to the total number of pixels for the considered bacterium: influence of the culture age and centrifugation process (see main text for more detailed information).

Figure 3:

AFM images on the top of bacteria at a high magnification for non-centrifuged and centrifuged samples at 3h culture: (0.4 x 0.4 µm)² scanning areas; figures 3.a: height images (a quadratic fit was removed from the raw data to enhance local roughness); figures 3.b: Young's modulus images; figures 3.c: histograms of figures 3.b.

Figure 4:

Real-time AFM imaging of *S. aureus* bacterial division on a non-centrifuged sample at 3h culture. (A) Height image (1.5µm)² ; (B) (400nm)² zoom at the black square in image 4.a; (C) Cross-sections along the red dashed line in figure 4.a illustrating the increase over time of the length and depth at the trench (bold red arrow in figure 4.a) resulting from the cell division process. (D) Variation of the gap length (see figure 4.c for definition) with time.

Figure 5:

AFM multiparametric imaging reveals the nanoscale structure and mechanical properties of *S. aureus* cell surface depending on its growth phase for non-centrifuged samples. For each condition (3-16-18-24h of growth phase), are represented: (A) height images at low magnification (scale bar: 1µm), (B.1-4) (0.4)² µm² height images corrected from curvature radii at respective black squares drawn at the top of bacteria in figures 5.a; (B.5) (0.3)² µm² height images corrected from curvature radii at red square drawn in figure 5.a.4 corresponding to the globular cluster structure. Two examples of such structure are drawn at the heads of red arrows; (C) corresponding elasticity maps for height images in figures 5.b and (D) the corresponding Young's moduli distribution histograms. Similar results were obtained for at least four cells from different cultures. Blue arrows in figures 5.b-c.2 indicate the holes (the "the *murosomes*[7]-mediated punching of holes into the peripheral wall for cell"[7]) formed at the bacterial surface. Blue rectangle in images 5.b-c.3 indicates the 'zipper-like' pattern (see main text for more details).

Figure 6:

Secretion of the extracellular polymeric material: white plume in figure 6.a and 6.b. Figure 6.b is a magnification of the zone in the black square in figure 6.a. The successive positions of the diffusion front are visible in figure 6.c. From that it is possible to estimate a diffusion/secretion coefficient for the extracellular material secreted by *S. aureus* bacteria (figure 6.d).

Figure 7:

Schematic representation of the proposed mechanism (the blue background at the right schematics visualizes the soft extracellular polymeric substance secreted by the bacteria).

Figure 1

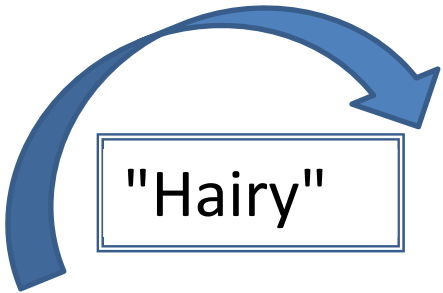

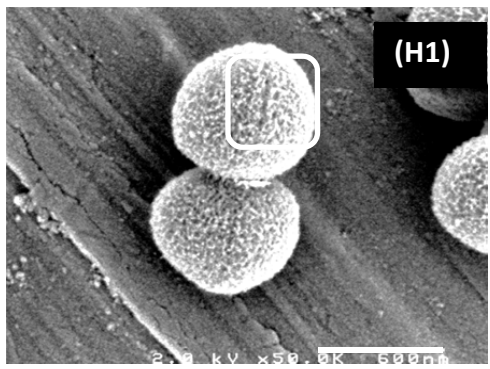
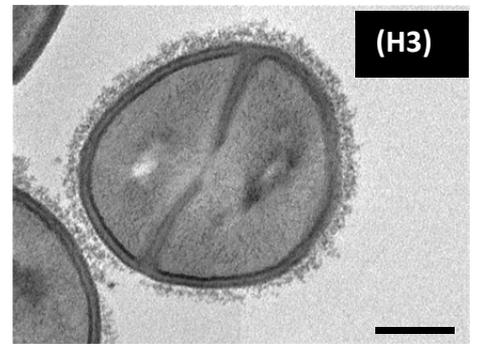
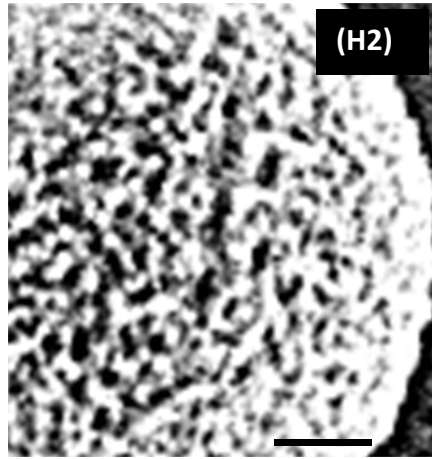
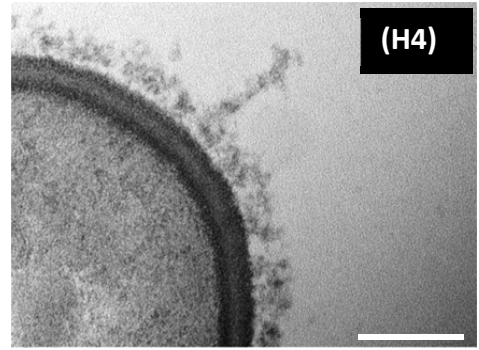
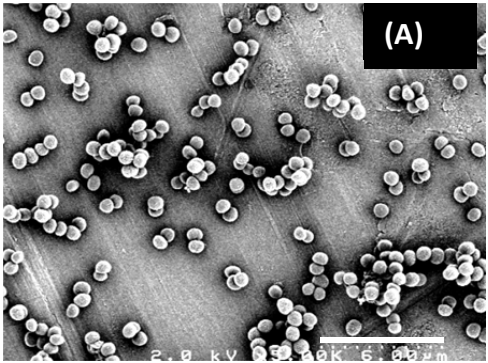
"Hairy"
"Bald"
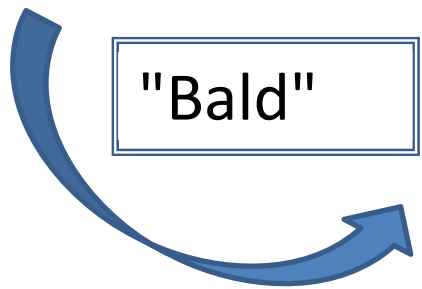
SEM
TEM
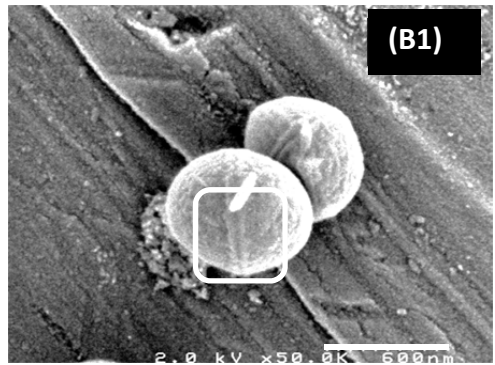
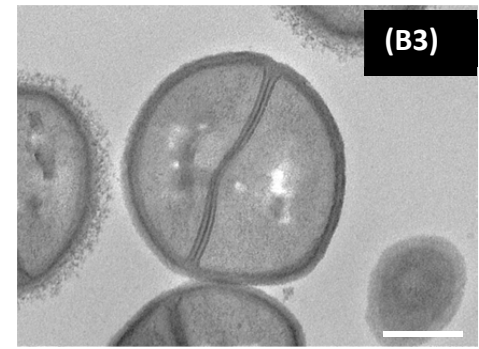
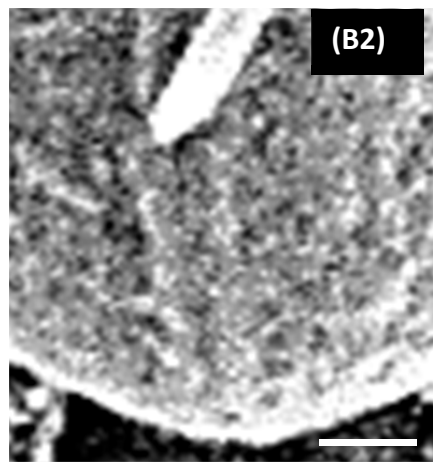
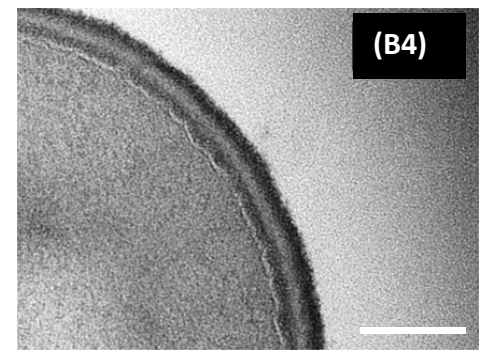

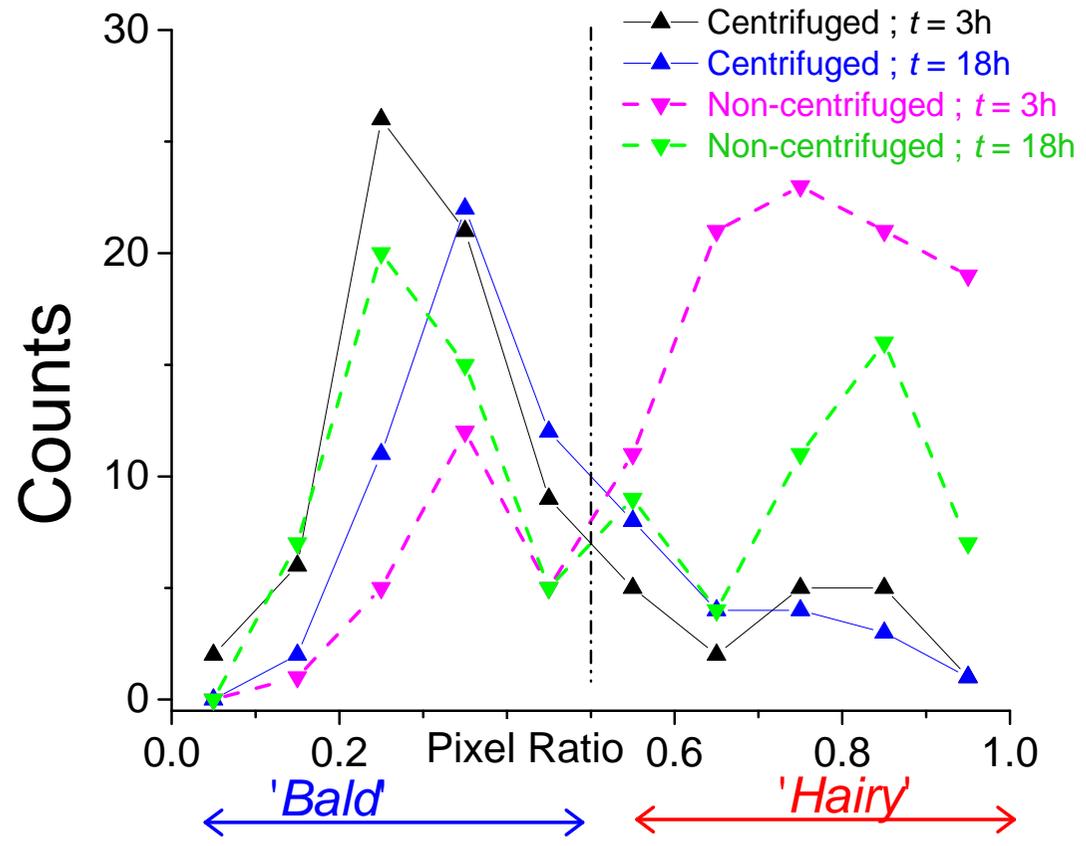

Figure 2.

| NON-CENTRIFUGED | CENTRIFUGED |
|---|---|
| 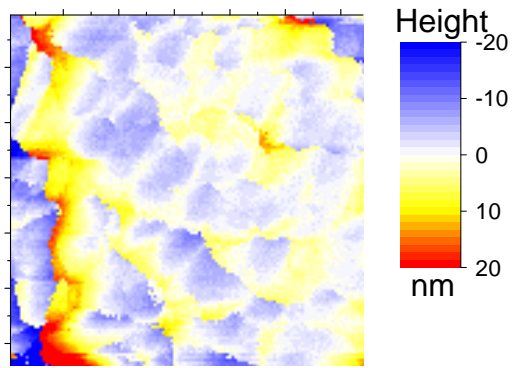 | 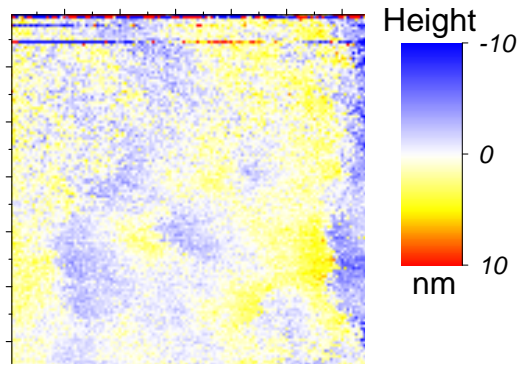 |
| Figure 3.a.1 | Figure 3.a.2 |
| 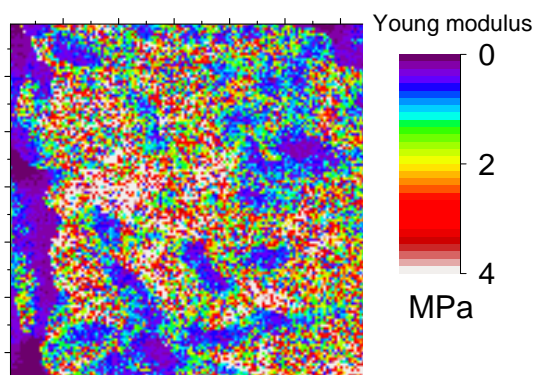 | 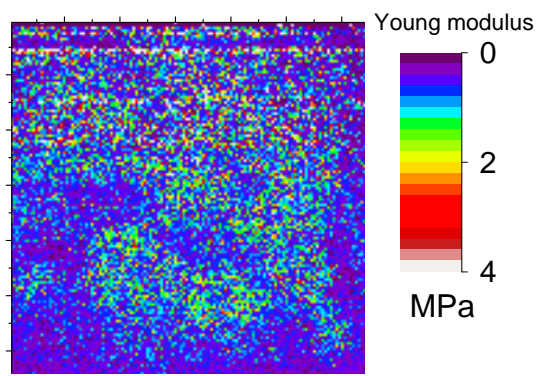 |
| Figure 3.b.1 | Figure 3.b.2 |
| 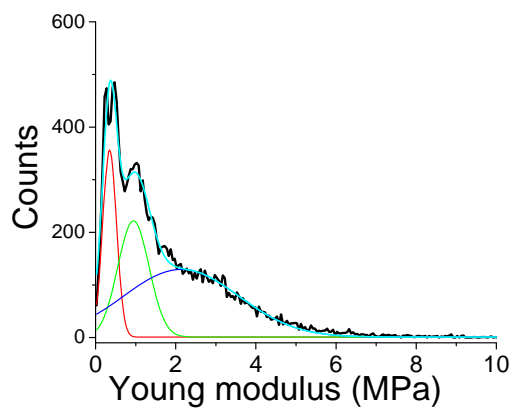 | 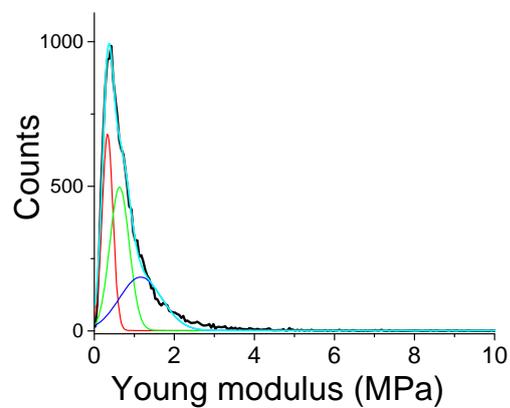 |
| Figure 3.c.1 | Figure 3.c.2 |

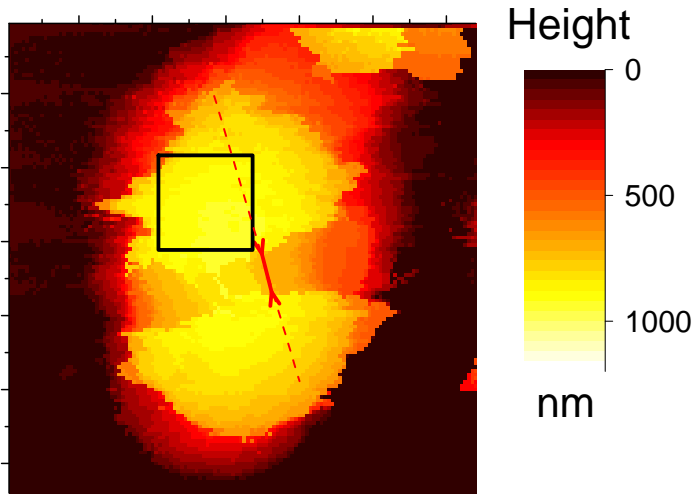

Figure 4.a

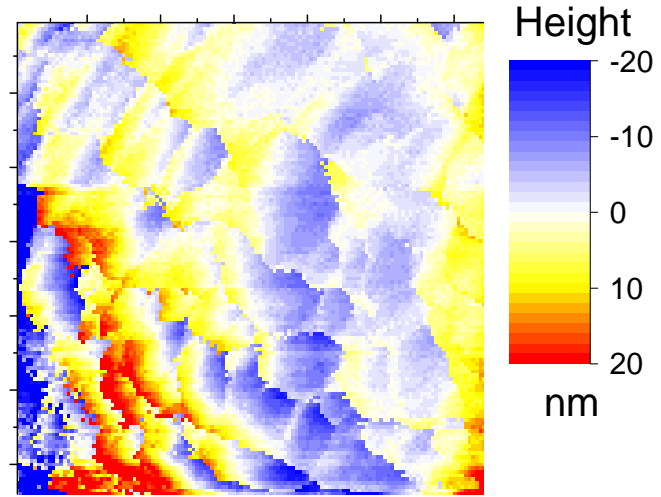

Figure 4.b

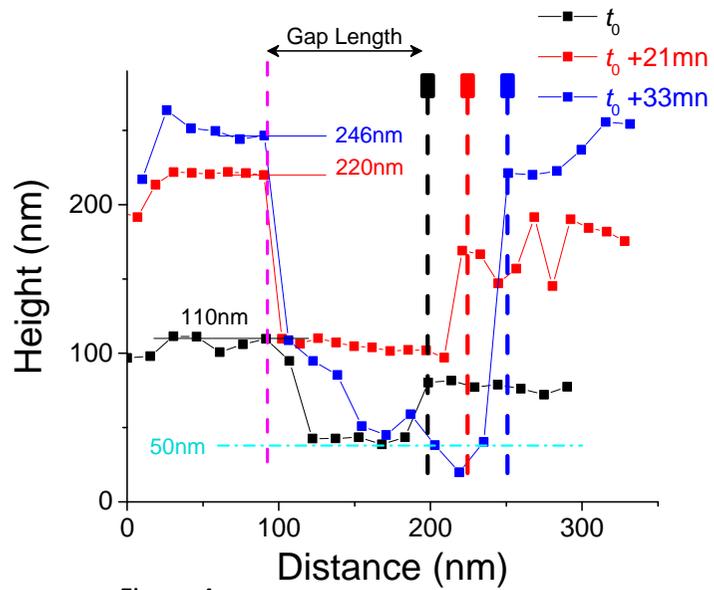

Figure 4.c

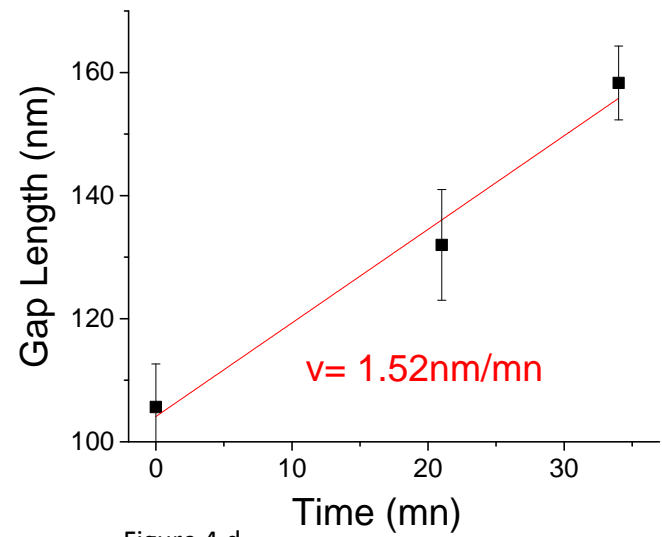

Figure 4.d

| Exponential Growth Phase (3h culture) | Early-stage Stationary Phase (16h culture) | Middle-stage Stationary Phase (18h culture) | Late-stage Stationary Phase (24h culture) | |
|---|---|---|---|---|
| 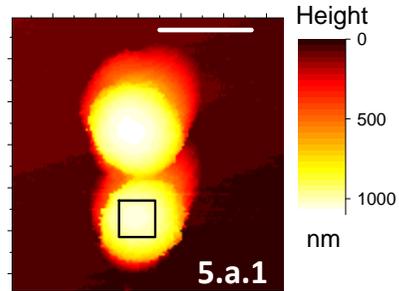 5.a.1 | 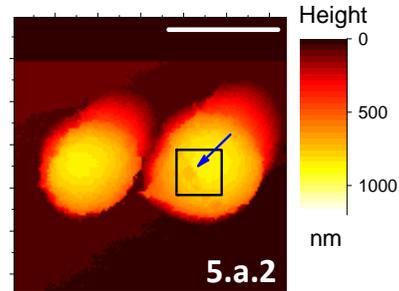 5.a.2 | 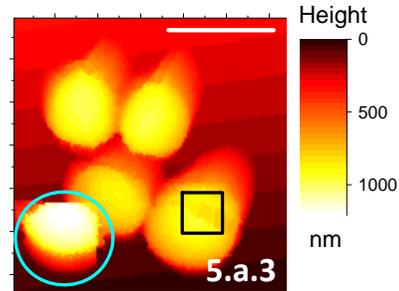 5.a.3 | 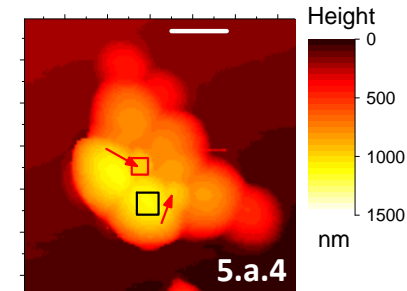 5.a.4 | |
| 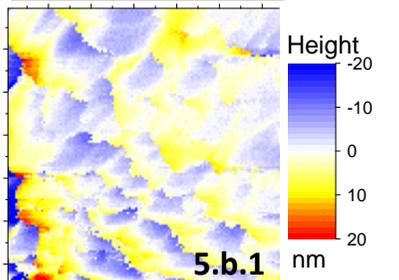 5.b.1 | 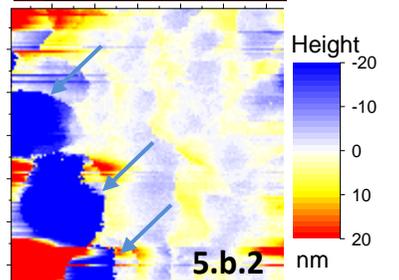 5.b.2 | 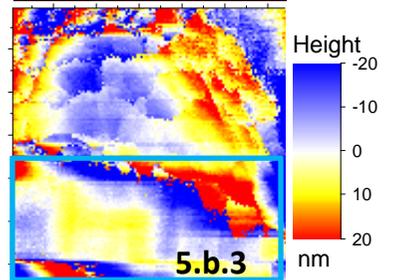 5.b.3 | 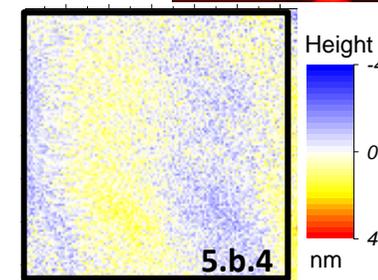 5.b.4 | 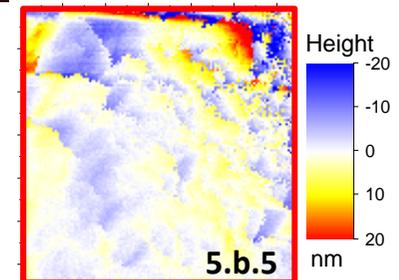 5.b.5 |
| 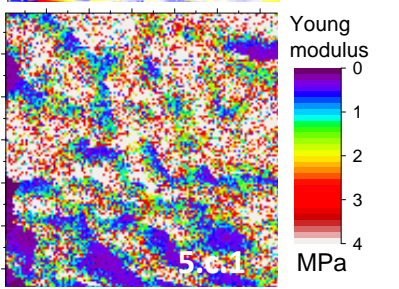 5.c.1 | 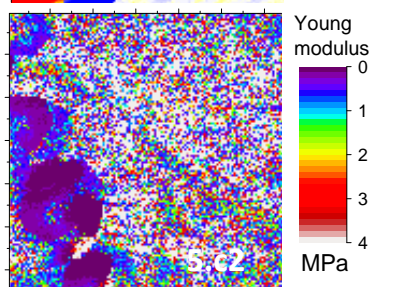 5.c.2 | 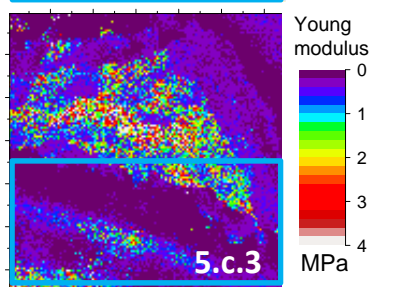 5.c.3 | 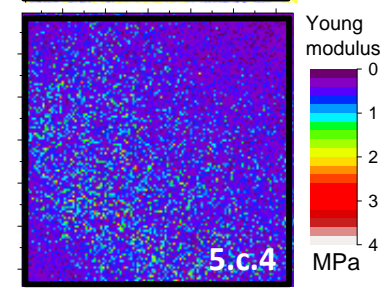 5.c.4 | 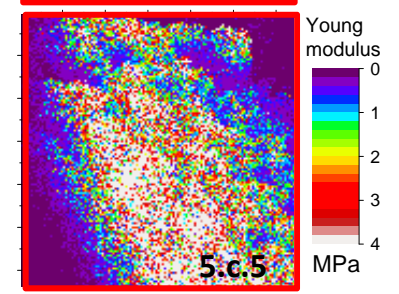 5.c.5 |
| 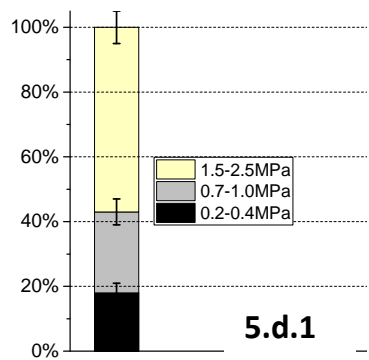 5.d.1 | 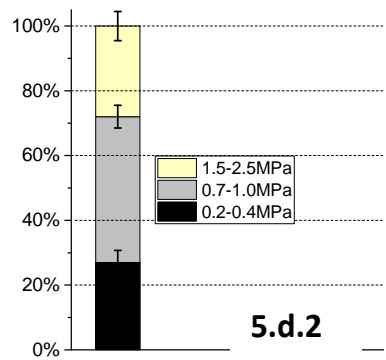 5.d.2 | 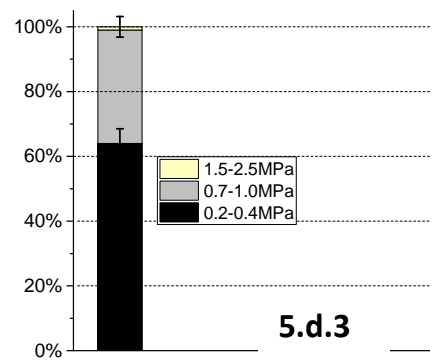 5.d.3 | 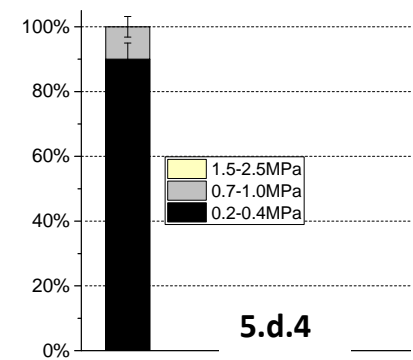 5.d.4 | 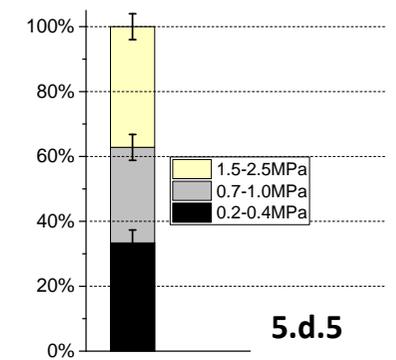 5.d.5 |

Figure 5

**Late-stage Stationary Phase**
**(24h culture)**

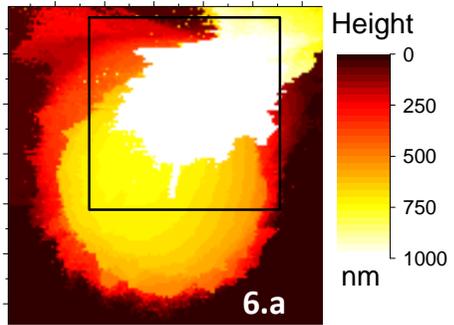 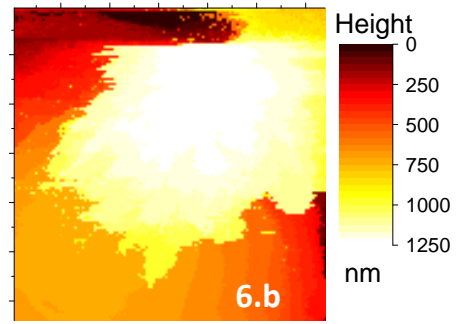 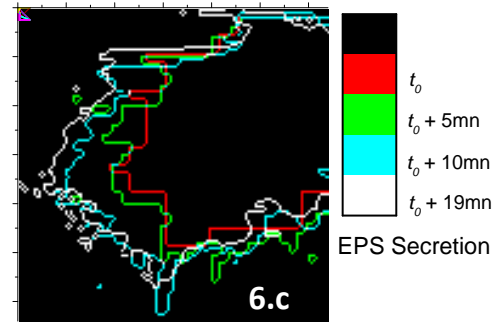 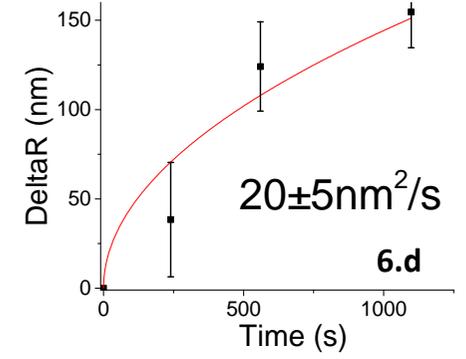

Figure 6

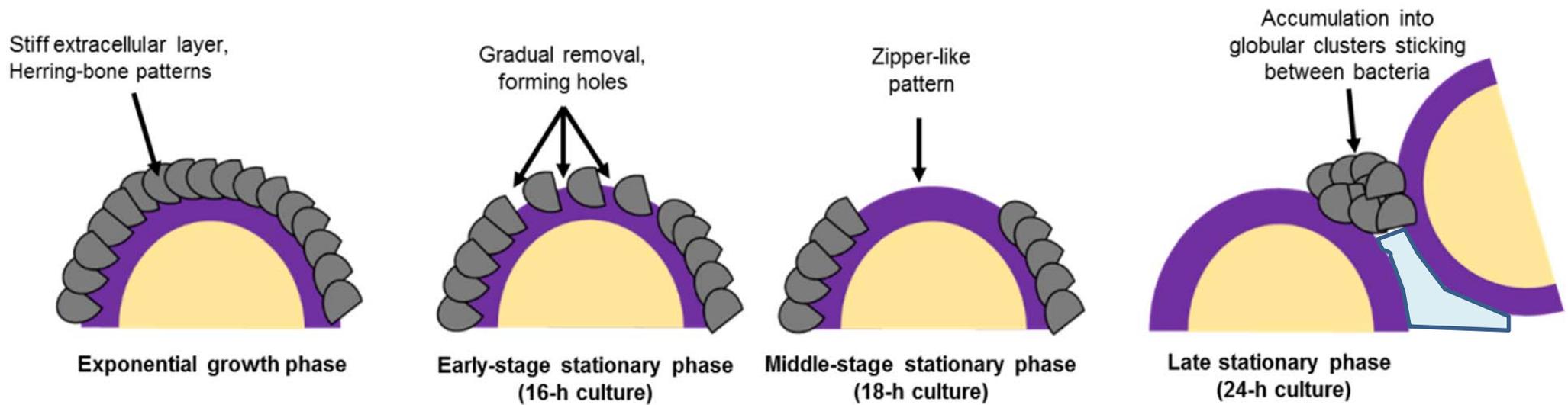

Figure 7

# Declaration of no conflict of interest

The authors of the paper entitled "*Direct observation of the cell-wall remodeling in adhering Staphylococcus aureus 27217: an AFM study supported by SEM and TEM*", submitted in the journal The Cell Suface (Elsevier) declare no competing interests.

December 19th 2018

# Supplementary Information

**Figure captions**

Figure SI1:

Principle of automatic detection of the "pixel-ratio" for each bacterium of the samples by a home-made Matlab program[1] constituted by the following steps:

1. the initial grayscale SEM image is converted to a binary one;

2. an on-screen first determination of the mean radius, $r_0$, of bacteria is performed;

3. the connected components in the binary image are detected and their edges determined;

4. these edges are approximated by circles with a variable radius the initial value of which was equal to the $r_0$, by using a circular Hough transform[2,3]. We checked the validity of such a process by verifying that all the cells were detected and that their final radius was in the range $r_0 \pm 10\%$;

5. successive dilation, erosion operations to increase the contrast of the cell surface structures for an automatic detection are done;

6. finally, the software automatically counts the ratio of the 'on' pixels, where these surface structures are were revealed, to the total number of pixels for each bacterium, the so-called "pixel-ratio".

Figure SI2:

SEM observation of *S. aureus* ATCC 27217 bacterial strain. Evidence of two types of self-adhering subpopulations: the so-called "*bald*" (figures SI2.b, SI2.d) and "*hairy*" cells (figures SI2.a, SI2.c). Figures SI2.c and SI2.d are numerical enlargements (with enhanced contrast) of images SI2.a and SI2.b (respectively) along the green squares. Scale bars: 500nm.

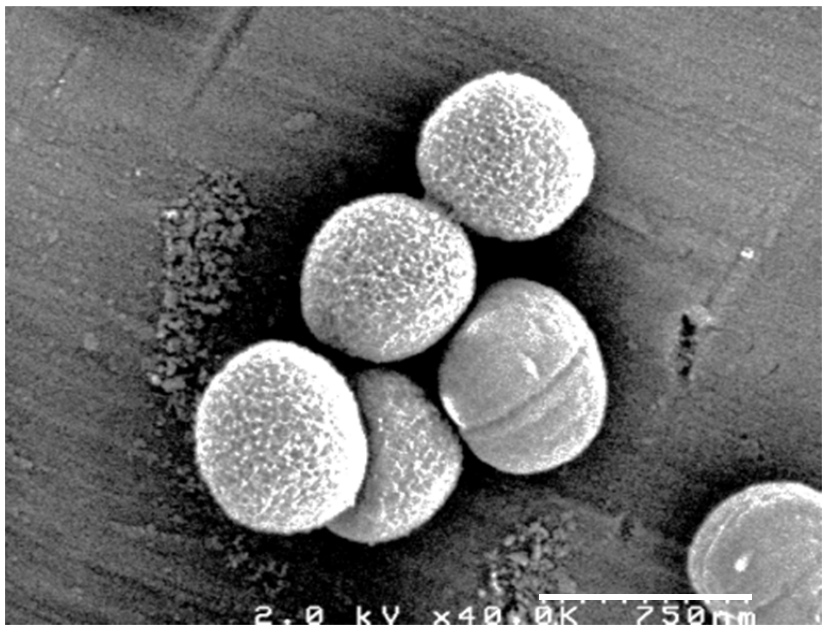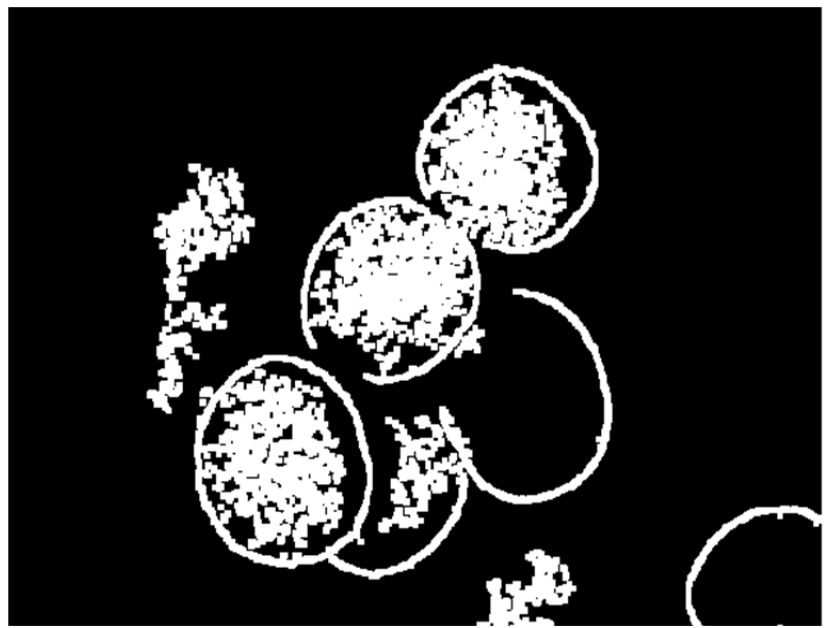

Figure SI1.a                                                                                       Figure SI1.b

| Non- Centrifuged | Centrifuged |
|---|---|
| 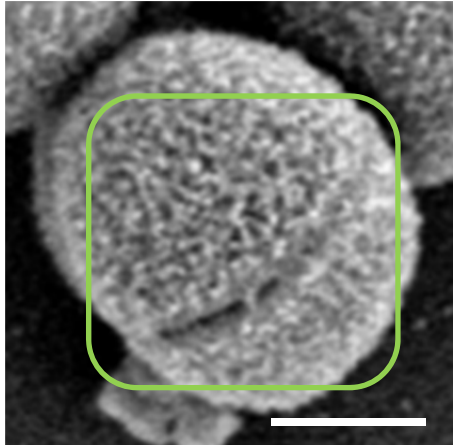 Figure SI2.a | 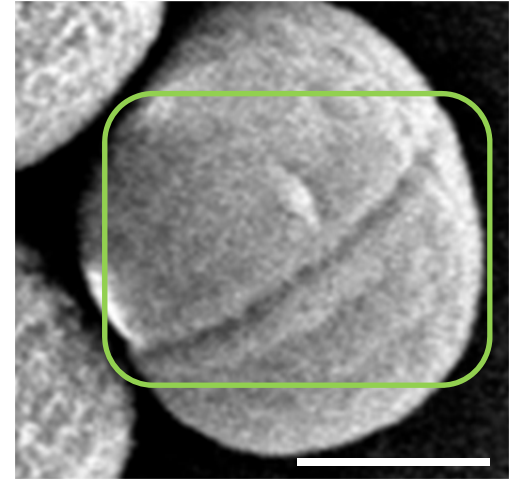 Figure SI2.b |
| 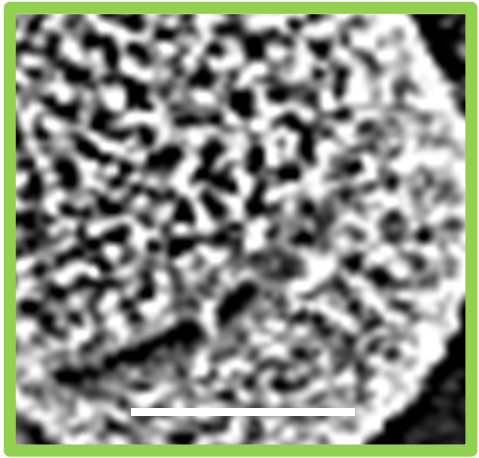 Figure SI2.c | 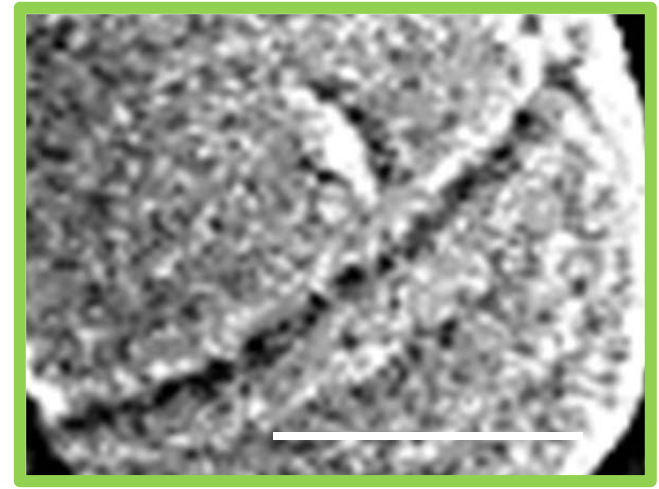 Figure SI2.d |